\documentclass [twocolumn,footinbib, bibnotes,showpacs,amssymb,
amsmath,aps,pra]{revtex4-1}
\usepackage[utf8]{inputenc}
\usepackage{bm}
\usepackage{graphics}
\usepackage{graphicx}
\usepackage{color}
\usepackage{amsmath}
\usepackage{amssymb}
\usepackage{url}

\begin{document}
\title{Entanglement and area laws in weakly correlated gaussian states}
\author{J.M. Matera, R. Rossignoli, N. Canosa}
\affiliation{Departamento de F\'{\i}sica-IFLP,
Fac.\ de Ciencias Exactas,
Universidad Nacional de La Plata, C.C. 67, La Plata (1900), Argentina}
\date{\today}

\begin{abstract}

We examine the evaluation of entanglement measures in weakly
correlated gaussian states. It is shown that they can be expressed
in terms of the singular values of a particular block of the
generalized contraction matrix. This result enables to obtain in a
simple way asymptotic expressions and related area laws for the
entanglement entropy of bipartitions in pure states, as well as
for the logarithmic negativity associated with bipartitions and
also pairs of arbitrary subsystems. As illustration, we consider
different types of contiguous and noncontiguous blocks in two
dimensional lattices. Exact asymptotic expressions are provided
for both first neighbor and full range couplings, which lead in
the first case to area laws depending on the orientation and
separation of the blocks.

\end{abstract}
\maketitle

\section{Introduction}

Entanglement is a valuable resource that plays a key role in
quantum information processing and transmission based on qubits
\cite{NC.00,BB.93,JL.03,RB.01} or on continuous variable systems
\cite{WP.12}. It has also provided new insights into the role of
quantum correlations in the critical behavior of many-body quantum
systems \cite{OA.02,ON.02,VLRK.03,AFOV.08,ECP.10}. Nonetheless,
the evaluation of genuine quantum correlations for a given state
of a many-body system is in general a difficult task. On the one
hand, rigorous computable entanglement measures exist just for
pure states, where  the entanglement entropy, i.e. the entropy of
the reduced state of a subsystem, provides the basic measure of
bipartite entanglement \cite{BBPS.96}. In the case of mixed
states, rigorous measures like the  entanglement of formation
\cite{BVSW.96}, which is the convex-roof extension of the previous
measure \cite{RC.03}, involve a minimization over a very high
dimensional space of parameters and are therefore not directly
computable. This has turned the attention to the
negativity \cite{VW.02}, or equivalently, the logarithmic
negativity \cite{VW.02,Pl.05}, which quantifies the violation of
the positive partial transpose separability criterion by entangled
states and is a bipartite entanglement monotone \cite{VW.02},
computable in principle for any bipartition in any pure or mixed
state. Nevertheless, the accurate evaluation of these quantities
demands a deep knowledge of the many body state, which requires in
general an amount of information which increases exponentially
with the system size. This fact limits the possibility of closed
evaluations to states characterized by a manageable number of
parameters \cite{AFOV.08}.

A prime example of such states are the {\it gaussian states,}
i.e., ground or thermal states of stable gapped Hamiltonians
quadratic in boson operators, or equivalently generalized
coordinates and momenta \cite{WP.12,HSH.1999,RS.00}. For such
states, of crucial importance for continuous variable quantum
information \cite{WP.12}, the entanglement entropy of bipartitions
of pure states and the negativity between arbitrary subsystems in
pure or mixed states can be evaluated in terms of the elements of
the covariance matrix \cite{AEPW.02,CEPD.06,CEP.07,ASI.04,AI.08},
i.e., of the generalized  contraction matrix of pairs of boson
operators \cite{RS.80,MRC.10}. However, even in this scenario, the
extraction of analytic expressions for these quantities for
arbitrary subsystems is in general not straightforward
\cite{AEPW.02,CEPD.06,MRPR.09}.

The aim of this work is to discuss the evaluation of the previous
measures in weakly correlated gaussian states, such as typical
ground states of gapped hamiltonians, which can be characterized
by excitations over a product state. Gaussian states are usually
described in terms of the phase space formalism \cite{WP.12},
which allows to connect their entanglement properties with
correlations in phase space. Here we will consider a different
approach, based on the Fock representation, which provides an
equivalent yet in many cases more clear way to evaluate and
represent entanglement measures \cite{MRC.10}. We will show that
the entanglement entropy and negativity can be expressed in terms
of the singular values of sub-blocks of basic contraction
matrices, which can be evaluated analytically in the perturbative
limit for some typical couplings. The formalism also allows the
straightforward derivation of area laws
\cite{ECP.10,CEPD.06,CC.04,JK.04,LORV.05} for these quantities.
The emergent area laws for the entanglement entropy and negativity
are different, i.e., they depend on distinct measures of the
boundary size, and are affected by the orientation and separation
of the subsystems. Let us also remark that the ground state of
weakly interacting spin systems can also be described by gaussian
states through different approximate bosonization techniques
\cite{RS.80,MRC.10,RCM.11}, entailing that the scope of the
present scheme is quite general.

The formalism is described in section II, while section III
considers its application to specific systems, essentially ground
states of two dimensional lattices with short range couplings,
although  the full range case is also considered. The present
scheme allows to easily obtain exact analytic asymptotic
expressions for the entanglement entropy and logarithmic
negativity of different types of bipartitions and block pairs,
both contiguous and separated, which will be compared with exact
numerical results. They clearly show the emergence of precise area
laws. Conclusions are finally drawn in IV.  We also include
appendices containing the details of the perturbative expansion
for the symplectic eigenvalue problem and the evaluation of
singular values.

 \section{Formalism}
\subsection{Entanglement entropy and negativity in Gaussian states}
The class of Gaussian states in a bosonic system can be defined as
those states of the form
\begin{equation}
\label{eq:defGaussStates} \rho={\textstyle\frac{1}{{\rm Tr}
\exp(-\beta H)}}T(\alpha)\exp(-\beta H)T^{\dagger}(\alpha)\,,
\end{equation}
where $H$ is a positive definite quadratic form on boson operators $b_i$,
$b^\dagger_i$ ($[b_i,b^\dagger_j]=\delta_{ij}$, $[b_i,b_j]=0$),
\begin{eqnarray}
\label{eq:H1}
H&=&\sum_{i,j} (\lambda_i\delta_{ij}-\Delta^+_{ij})
(b^\dagger_i b_j+{\textstyle\frac{1}{2}}\delta_{ij})-{\textstyle\frac{1}{2}}
(\Delta^-_{ij}b_ib_j+\bar{\Delta}^-_{ij}b^\dagger_jb^\dagger_i)\nonumber
\\&=&{\textstyle\frac{1}{2}}{\cal Z}^\dagger  {\cal H} {\cal Z}\,,\;\;\;
  \label{eq:defQuadrBF}
  {\cal H}=\left(\begin{array}{c c}
\Lambda -\Delta^{+} & -\Delta^{-}\\
-\bar{\Delta}^{-} & \Lambda -\bar{\Delta}^{+}
\end{array}\right)\,,
\end{eqnarray}
with ${\cal Z}=\left(^{b}_{b^{\dagger}}\right)$, and
$T(\alpha)=\prod_i\exp(\bar{\alpha}_i b_i-\alpha_i b^{\dagger}_i)$
is a displacement operator
($T(\alpha)b_iT^\dagger(\alpha)=b_i+\alpha_i$). In
(\ref{eq:defQuadrBF}),  $\Lambda$ is the diagonal matrix
of local bare energies $\lambda_i$
and $\Delta_{ij}^{\pm}$ are the coupling strengths between pairs of
different bosons ($\Delta_{ij}^+=\bar{\Delta}_{ji}^+$,
$\Delta_{ij}^-=\Delta_{ji}^-$). In the pure state limit
$\beta\rightarrow \infty$, $\rho\rightarrow T(\alpha)|0\rangle
\langle 0|T^\dagger(\alpha)$, with $|0\rangle$ the ground state of
$H$. The displacements $\alpha_i$ can be
taken into account by local shifts  $b_i\rightarrow b_i-\alpha_i$,
so that in what follows we will set $\alpha_i=0$, such that
$\langle b_i\rangle_\rho\equiv{\rm Tr}\,\rho\,b_i=0$ $\forall$
$i$.

The key property of these states is that by means of Wick's
theorem \cite{RS.80}, the expectation value of any bosonic
operator (and hence $\rho$) is fully determined by the
displacements  $\alpha_i$  and the generalized contraction
matrix\cite{MRC.10,RS.80}
\begin{eqnarray}
  {\cal D}&=&\langle{\cal Z}{\cal Z}^\dagger\rangle-{\cal M}
  \label{defD}
= \left(
\begin{array}{c c}
F^{+}&  F^{-}\\
\bar{F}^{-}&\bm{1}+\bar{F}^{+}
\end{array}
\right)\,,\nonumber\\
  F_{ij}^{+}&=&\langle b_j^\dagger b_i\rangle_\rho\,,\;\;
F_{ij}^{-}=\langle b_ib_j\rangle_\rho\,, \label{eq:BlockCovMatrix}
\end{eqnarray}
where ${\cal M}={\cal Z}{\cal Z}^{\dagger}-[({\cal
Z}^\dagger)^t{\cal Z}^t]^t=(^{\bm{1}\;\;\;0}_{0\,-\bm{1}})$ is the
\emph{symplectic metric} and $F^+_{ij}=\bar{F}_{ji}^+$,
$F^-_{ij}=F^-_{ji}$.

We may diagonalize ${\cal D}$ or ${\cal H}$  by means of a
symplectic transformation ${\cal W}$ satisfying ${\cal W}^\dagger
{\cal M}{\cal W}={\cal M}$, corresponding to a Bogoliubov
transformation ${\cal Z}={\cal W}{\cal Z}'$ to boson operators
${\cal Z'}=(^{b'}_{{b'}^\dagger})$, such that ${\cal D}={\cal
W}{\cal D}'{\cal W}^\dagger$, with ${\cal D}'$ diagonal
(${F'}^+_{\alpha\alpha'}=f_\alpha\delta_{\alpha\alpha'}$,
${F'}^{-}=0$). This leads to the standard diagonalization of the
matrix ${\cal D}{\cal M}$ (as ${\cal W}{\cal D}{\cal M}{\cal
W}^{-1}={\cal D}'{\cal M}$). The matrix ${\cal W}$ can be written
in block form as
\begin{equation}
  \label{eq:WBlocks}
  {\cal W}=\left(
\begin{array}{cc}
{U}& {V} \\
\bar{V}&\bar{U}
\end{array}
\right)\,,
\end{equation}
where ${U}$ and ${V}$ should satisfy
\begin{subequations}
\label{eq:relUV}
\begin{eqnarray}
 {U}^{\dagger}{U}-{V}^{\rm t}\bar{V}&=&\bm{1} ={U}
 {U}^{\dagger}-{V}{V}^{\dagger}
  \label{eq:relUVa}\\
 {U}^{\dagger}{V}-{V}^{\rm t}\bar{U} &=&0={U}{V}^{\rm t}
 -{V}{U}^t \,.\label{eq:relUVb}
\end {eqnarray}
\end{subequations}
The blocks $F^{\pm}$ of the contraction matrix
acquire then a simple form in terms of $U$, $V$ and the diagonal block
${F'}^+$:\begin{subequations}
\label{FUV}
\begin{eqnarray}
  {F}^{-}&=&VU^t+V{F'}^+U^t+U{F'}^+V^t\\
  {F}^{+}&=&VV^{\dagger}+V{F'}^+V^{\dagger}+U{F'}^+U^{\dagger}\,.
\end{eqnarray}
\end{subequations}
For a {\it pure state}, ${F'}^+=0$ and Eqs.\ (\ref{FUV}) lead to
$F^-=VU^t$, $F^+=VV^\dagger$, implying
\begin{equation}F^-\bar{F}^-=F^++{F^+}^2\label{Fex}\,.\end{equation}

For such states, the \emph{entanglement} between any subsystem
${\cal A}$ and its complement $\bar{\cal A}$ can be measured
through the von Newmann entropy of any of the reduced  states:
\begin{equation}
{\cal E}_{{\cal A},\overline{\cal A}}=S(\rho_{\cal
A})=S(\rho_{\overline{\cal A}})\,,
 \label{SA}\end{equation}
where $S(\rho)=-{\rm Tr}\,\rho \log \rho$. In a Gaussian state,
the validity of Wick's theorem\cite{RS.80} implies that the state
of any subsystem ${\cal A}$ is also Gaussian and hence fully
characterized by the corresponding contraction matrix ${\cal
D}_A$,  which is just the sub-block of ${\cal D}$ with indexes
belonging to  ${\cal A}$:
\begin{eqnarray}
  {\cal D}_{\cal A}&=
  \label{defDA}
\left(
\begin{array}{c c}
F^{+}_{\cal A}&  F^{-}_{\cal A}\\
\bar{F}^{-}_{\cal A}&\bm{1}+\bar{F}^{+}_{\cal A}
\end{array}
\right)\,.
\end{eqnarray}
The von Newmann entropy (\ref{SA}) can then be expressed in terms
of the symplectic eigenvalues $f^{\cal A}_\alpha$ of ${\cal
D}_{\cal A}$ as
\begin{equation}
S(\rho_{\cal A})=\sum_\alpha h(f^{\cal A}_\alpha)\,,\;\;h(x)=-x
\log x+(1+x)\log(1+x)\,.\label{hx}
\end{equation}

In the case of a mixed state or for pairs of non-complementary
subsystems ${\cal B,C}$, the subsystem entropy is no longer a
measure of quantum correlations. Instead, a well known computable
entanglement monotone for such systems is the {\it negativity}
$N_{\cal B,C}$ \cite{VW.02}, which is just the sum of the negative
eigenvalues of the \emph{partial transpose} $\rho_{\cal B\cal
C}^{t_{\cal B}}$.  An associated quantity is  the
\emph{logarithmic negativity}
\begin{equation}
{\cal E}^{\cal N}_{\cal B,C}=\log (1+2N_{\cal B,C})
=\log ||\rho_{\cal B\cal C}^{\rm t_{\cal B}} ||_1\,,\label{LN}
\end{equation}
where $||A||_1={\rm tr}\,\sqrt{A^{\dagger}A}$ denotes the
\emph{trace norm}.  For a gaussian state,
$||\rho_{\cal B\cal C}^{\rm t_{\cal B}} ||_1$ can be expressed in
terms of the \emph{negative symplectic eigenvalues}
$\tilde{f}^{\cal B,C}_\alpha$ of the contraction matrix
$\tilde{\cal D}_{\cal BC}$ determined by
$\rho_{\cal{B}\cal{C}}^{{\rm t}_{\cal B}}$, with blocks
\begin{equation}
  \tilde{F}^{\pm}_{\cal BC}=\left(\begin{array}{c c}
\bar{F}_{\cal B}^{\pm} & \bar{F}_{\cal B,C}^{\mp}\\
F_{\cal C,B}^{\mp} & F_{\cal C}^{\pm}
\end{array}\right)\,,\label{Ft}
\end{equation}
where  $F^\pm_{\cal B,C}$ denotes the matrix of elements
$F^\pm_{ij}$ with $i\in{\cal B}$ and $j\in{\cal C}$, and
$F^\pm_{\cal B}\equiv F^\pm_{\cal B,B}$. Eq.\ (\ref{LN}) then
becomes
\begin{equation}
  {\cal E}^{\cal N}_{{\cal B},{\cal C}}=
  \sum_{\tilde{f}^{\cal B,C}_\alpha<0} g(\tilde{f}^{\cal B,C}_\alpha)\,,\;\;
g(x)=-\log(1+2 x)\,.\label{lneg}
\end{equation}
We notice that $\tilde{f}^{\cal B,C}_{\alpha}\geq -1/2$
\cite{MRC.10}. In the case where ${\cal B}={\cal A}$ and ${\cal
C}={\cal \bar{A}}$, Eq.\ (\ref{lneg}) can be expressed in terms of
the symplectic eigenvalues $f_\alpha ^{\cal A}$ of ${\cal D}_{\cal
A}$ as \cite{MRC.10}
\begin{equation}
  {\cal E}^{\cal N}_{\cal A,\bar{A}}=2\sum_{\alpha} \log(\sqrt{f_\alpha^
 {\cal A}}+\sqrt{1+f_\alpha^{\cal A}})\,,\label{lnegx}
\end{equation}
which, like Eq.\ (\ref{SA}), is again a concave function of the
$f^{\cal A}_\alpha$.

\subsection{Weakly correlated pure Gaussian states}
We consider now the case of weakly correlated pure Gaussian
states, i.e., states in which the local symplectic eigenvalues
(those corresponding to a single mode ${\cal A}=i$)
\begin{equation}
  f_i={\textstyle\sqrt{(\frac{1}{2}+F^+_{ii})^2-
  |F_{ii}^-|^2}-\frac{1}{2}}\,, \label{eq:ssexactf}
\end{equation}
satisfy $f_{i}\ll 1$ $\forall$ $i$, such that each mode is weakly
entangled with the rest of the system. In the local basis where
$F^-_{ii}=0$ (this implies replacing $b_i\rightarrow
u_ib_i-e^{i\phi}v_ib^\dagger_i$, with
$u_i,v_i=\sqrt{\frac{F^+_{ii}+1/2\pm(f_i+1/2)}{2f_i+1}}$ and
$\phi$  the phase of $F^-_{ii}$),  $f_i=F^+_{ii}$ and weak
coupling implies, together with the positivity of ${\cal D}$ and
${F}^+$, that $|F^+_{ij}|\leq \sqrt{f_i f_j}\ll 1$,
$|F^{-}_{ij}|^2\leq {\rm Min}[f_i,f_j]+f_if_j\ll 1$ $\forall$
$i,j$. In this limit, Eqs.\ (\ref{FUV})--(\ref{Fex}) then lead,
neglecting terms $\propto (F^-\bar{F}^-)^2$,  to
\begin{equation}
F^+\approx F^-\bar{F}^-\,,\label{F1}
\end{equation}
which for a subsystem ${\cal A}$ implies
\begin{equation}
F^+_{\cal A}\approx F^-_{\cal A}\bar{F}^-_{\cal A}+
F^-_{\cal A,\bar{A}}\bar{F}^-_{\cal \bar{A},A}\,.
\label{F2}
\end{equation}
Using Eq.\ (\ref{F2}) and the results of Appendix
\ref{app:perturb}, the symplectic eigenvalues of ${\cal D}_{\cal
A}$  will then agree at this order with the standard eigenvalues
of the matrix
\begin{equation}
  F^+_{\cal A}-F^-_{\cal A}\bar{F}^-_{\cal A}\approx
   F^-_{\cal A,\bar{A}}\bar{F}^-_{\cal \bar{A},A}\,,\label{eq:FplusWCS}
\end{equation}
which are just the square of the {\it singular values}
$\sigma_\alpha^{\cal A,\bar{A}}$ of $F^-_{\cal A,\bar{A}}$ (see
Appendix \ref{S}). We then obtain, at this order,
\begin{equation}
f^{\cal A}_\alpha\approx(\sigma_\alpha^{\cal A,\bar{A}})^2\,.\label{Sw}
\end{equation}
Entanglement depends at this level just on the $F^-$ contractions
between ${\cal A}$ and ${\cal \bar{A}}$. For instance, in the case
of a single site $i$, Eq. (\ref{Sw}) implies $f_i\approx
\sigma^2_{i,\bar{i}}=\sum_{j\neq i}|F^-_{ij}|^2$.

In this regime we may just set $h(x)\approx -x\log_2 (x/e)$ in
Eq.\ (\ref{hx}), such that the entanglement entropy becomes
\begin{equation}
{\cal E}_{\cal A,\bar{A}}\approx -\sum_\alpha (\sigma_\alpha^{\cal A,\bar{A}})^2
\log[(\sigma_\alpha^{\cal A,\bar{A}})^2/e]\,.\label{Sp}\end{equation}

Considering now the negativity, in the present regime the
symplectic eigenvalues of $\tilde{\cal D}_{\cal BC}$ will be given
at leading order by the eigenvalues of (see Appendix
\ref{app:perturb})
\begin{eqnarray}
\tilde{F}^+_{\cal BC}-\tilde{F}^-_{\cal BC}\bar{\tilde{F}}^-_{\cal BC}&\approx&
\left(\begin{array}{cc}
\bar{G}_{\cal B}&\bar{F}^-_{\cal B,C}\\F^-_{\cal C,B}&G_{\cal C}
\end{array}\right)\,,\label{eqneg0}\end{eqnarray}
where, for  ${\cal S}={\cal B}$ or ${\cal C}$,
\begin{equation}
G_{\cal S}=F^+_{\cal S}-{F}^-_{\cal S}\bar{F}^-_{\cal S}\,.\label{G1}
\end{equation}
For pure global states, Eq.\ (\ref{eq:FplusWCS}) leads to $G_{\cal
S}\approx{F}^-_{\cal S,\bar{S}}\bar{F}^-_{\cal\bar{S},S}$,
indicating that ${\cal G}_{\cal S}$ takes into account the
correlations with the environment of ${\cal S}$. Up to first order
in $F^-_{\cal B,C}$, we may neglect its second order effect in
$G_{\cal \bar{B}}$ and $G_{\cal C}$ in (\ref{eqneg0}), such that
\begin{equation} G_{\cal B}\approx F^-_{\cal B,\overline{BC}}
\bar{F}^-_{\cal \overline{BC},B},\;\;G_{\cal C}\approx F^-_{\cal
C,\overline{BC}}\bar{F}^-_{\cal
 \overline{BC},C}\,,\label{G}\end{equation}
depend just on the correlation with the environment of ${\cal B C}$. If
the sites of ${\cal B}$ and ${\cal C}$ correlated with each other
have correlations of the same order (or less) with $\overline{\cal
BC}$, (i.e. $||F^-_{\cal B,\overline{BC}}||_\infty$ and
$||F^-_{\cal C,\overline{BC}}||_\infty$  of the same order as
$||F^-_{\cal B,C}||_\infty$, at least for the subsets of ${\cal
B}$ and ${\cal C}$ mutually correlated) we can directly neglect
$G_{\cal B}$ and $G_{\cal C}$ in Eq.\ (\ref{eqneg0}) at order
$||F^-_{\cal B,C}||_\infty$. The negative symplectic eigenvalues
of $\tilde{\cal D}_{\cal B,C}$ will then be given by minus the
{\it singular values} $\sigma^{\cal B,C}_\alpha$ of $F^-_{\cal
B,C}$ (see Appendix \ref{S}):
\begin{equation}
\tilde{f}_{\alpha}^{\cal B,C}\approx-\sigma_\alpha^{\cal B,C}\,,
\label{ftap}
\end{equation}
which depend again just on the $F^-$ contractions between ${\cal
B}$ and ${\cal C}$. For instance, this is the case of contiguous
blocks in a scenario of short range couplings, and also that where
${\cal C}$ is the complement of ${\cal B}$ (${\cal C}={\cal
\bar{B}}$).

In the general case, however, the whole matrix
(\ref{eqneg0}) should be diagonalized. First order corrections lead to
\begin{equation}
\tilde{f}_\alpha^{\cal B,C}\approx-\sigma_{\alpha}^{\cal B,C}+
[(\bar{G}_{\cal B})_{\alpha\alpha}+(G_{\cal C})_{\alpha\alpha}]/2\,,
\label{Gw}
\end{equation}
where $(\bar{G}_{\cal
B})_{\alpha\alpha}=U_\alpha^\dagger\bar{G}_{\cal B}U_\alpha$,
$(G_{\cal C})_{\alpha\alpha}=V_\alpha^\dagger G_{\cal B}V_\alpha$
are the diagonal elements in the local basis of ${\cal B}$ and
${\cal C}$ where $({\cal F}^-_{\cal
B,C})_{\alpha\alpha'}=\sigma_{\alpha}^{\cal
B,C}\delta_{\alpha\alpha'}$ (see App.\ \ref{S}). As $G_{\cal B}$
and $G_{\cal C}$ are positive matrices in the approximations
(\ref{G1})--(\ref{G}),  negative eigenvalues can only arise if
$(G_{\cal B})_{\alpha\alpha}$ and $(G_{\cal C})_{\alpha\alpha}$
are not much larger than $\sigma^{\cal B,C}_\alpha$.  A sufficient
condition ensuring a  negative eigenvalue $\tilde{f}_\alpha^{\cal
B,C}$ of (\ref{eqneg0}) is
 \begin{equation}\sigma_{\alpha}^{\cal B,C}>
 \sqrt{(\bar{G}_{\cal B})_{\alpha\alpha} (G_{\cal C})_{\alpha\alpha}}
 \,.\label{Gex}\end{equation}

In the present regime, the logarithmic negativity can be obtained
setting $g(x)\approx -2\log(e) x$ in (\ref{lneg}), such that
\begin{equation} {\cal E}^{\cal N}_{\cal B,C}\approx
-2\log(e)\sum_{\tilde{f}^{\cal B,C}_\alpha<0}
\tilde{f}_\alpha^{\cal B,C}\label{Np}\,,
\end{equation}
i.e., ${\cal E}^{\cal N}_{\cal B,C}\propto||F^-_{\cal B,C}||_1$ in
the approximation (\ref{ftap}). For complementary subsystems
(${\cal C}=\bar{\cal B}=\bar{\cal A}$), it is verified that identity
between Eqs.\ (\ref{lnegx}) and (\ref{Np}) holds at leading order
in the approximation (\ref{Sw})
($\log(\sigma+\sqrt{1+\sigma^2})\approx \log(e)\sigma$).

\subsection{Ground state correlation matrix in the weakly interacting case}

A particular case of the previous results arises when we deal with
the ground state of a Hamiltonian of the form (\ref{eq:H1}). For
weak couplings $\Delta^{\pm} \ll \Lambda$, the diagonalizing
symplectic transformation ${\cal W}$ such that ${\cal W}^\dagger
{\cal H}{\cal W}=\Omega\oplus\Omega$, with
$\Omega_{\alpha\alpha'}=\delta_{\alpha\alpha'}\omega_\alpha$, can
be evaluated perturbatively. At leading order (see Appendix
\ref{app:perturb}) the block  $U$ in (\ref{eq:WBlocks}) is a
\emph{unitary} matrix that diagonalizes $\Lambda-\Delta^+$, while
\begin{equation}
  {V}_{i\beta}\approx   \sum_{\alpha} {U}_{i \alpha}
  \frac{  ({U}^{\dagger} \Delta^{-}\overline{ U})_{\alpha \beta}}
  {\omega_{\alpha}+\omega_{\beta}}\,.\label{eq:Vpert1}
\end{equation}
Note that, in contrast with the conventional perturbation theory,
a possible degeneracy in the local energies $\lambda_k$ will not
spoil this result if the system is stable
($\omega_\alpha>0$ $\forall$ $\alpha$). Notice, however, that
${U}$ can depart considerably from the identity if the $\lambda_k$
are degenerate.

If all local bare energies are nearly equal
($|\lambda_k-\lambda_j|\ll \lambda_k+\lambda_j\approx 2\lambda $),
and if energy corrections arising from $\Delta^+$ are neglected
($\omega_{\alpha} \approx \lambda$), Eq.  (\ref{eq:Vpert1})
reduces to $  {V}\approx   \frac{1}{2 \lambda}  \Delta^-\bar{U}$.
In such a case, Eq.\ (\ref{FUV})  leads to
\begin{equation}
  F^-\approx\frac{\Delta^-}{2\lambda}\,,
\label{ecFDelta}
\end{equation}
with $F^+$ given by Eq.\ (\ref{F1}). In this regime,
correlations are hence proportional to the pairing couplings
$\Delta^-$, decreasing as $\lambda^{-1}$ for increasing local
energies. Noteworthy,  the strength of the hopping interaction
$\Delta^+$ does not affect the ground state correlations at this
order. When non degenerate, it just affects $F^{\pm}$ dressing the
bare pairing interactions.

In the same way, for a common local bare energy $\lambda$,
inclusion of second order terms in the couplings leads to
\begin{equation}
 F^-\approx\frac{\Delta^-}{2\lambda} +
  \frac{\Delta^+\Delta^-+\Delta^-\bar{\Delta^+}}{4\lambda^2}\,.
\label{ecFDelta2}
\end{equation}
This expression is useful in the present scheme when the first
order term vanishes (modes $i,j$ unconnected by $\Delta^-$). As
the counter terms $G_{\cal B}$ and $C_{\cal C}$ in (\ref{Gw}) will
be of second order in $F^-$ (Eq.\ (\ref{G})), subsystems
unconnected by $\Delta^-$ but connected at second order through
Eq.\ (\ref{ecFDelta2}) may exhibit an $O(\Delta/\lambda)^2$
non-zero negativity if Eq.\ (\ref{Gex}) holds.

\subsection{Area laws}

The formulation of the area law for systems with local
interactions starts with the definition of a suitable measure for
the size of the  boundary $\partial {\cal A}$ of the  subsystem
${\cal A}$ \cite{CEPD.06,MRPR.09,ECP.10}.
An example of such measure is given by the number of
pairs of first neighbor modes, with one mode belonging to ${\cal
A}$ and the other to $\bar{\cal A}$. If we define the matrix $M$
with entries $M_{ij}=1$ if modes $i$ and  $j$ are first neighbors
and 0 otherwise, that measure can be written as
\begin{equation}
  |\partial {\cal A}|_{2}=\sum_{i \in {\cal A}} n^{\bar{\cal A}}_{i}
  ={\rm Tr}\,\left[ { M}_{{\cal A},\bar{\cal A}}
  { M}_{\bar{\cal A},{\cal A}}\right]=||M_{\cal A,\bar{A}}||_2^2
\label{eq:l2measure}
\end{equation}
where $n_i^{\bar{A}}=(M_{\cal A,\bar{A}}M_{\cal \bar{A},A})_{ii}$
is the number of first neighbors of mode $i$ in $\bar{\cal A}$. For
the ground state of a gapped  bosonic system with constant and
{\it isotropic} first neighbor interactions
$\Delta_{ij}^{\pm}=\frac{\Delta^{\pm}}{2}{ M}_{ij}$, Eq. \
(\ref{ecFDelta}) implies  $F_{{\cal A}\bar{\cal
A}}^{-}\approx\frac{\Delta^{-}}{4\lambda}{M}_{{\cal A}\bar{\cal
A}}$ and Eqs.\ (\ref{Sw})--(\ref{Sp}) lead then to
\begin{equation}
{\cal E}_{{\cal A},\bar{\cal A}}\propto |\partial {\cal A}|_2
 \,,\label{E2}\end{equation}
at leading order in $\Delta^-/\lambda$, which coincides exactly
with the result in \cite{CEPD.06} for non critical harmonic
systems.

The logarithmic negativity presents, however, a slightly different
behavior: for contiguous subsystems, the same procedure and Eqs.\
(\ref{ftap})--(\ref{Np}) lead to
\begin{equation}
{\cal E}^{{\cal N}}_{{\cal A},{\cal \bar{A}}}\propto
|\partial {\cal A}|_1\,,
\label{E1}
\end{equation}
where the boundary measure is now
 \begin{equation}
  |\partial {\cal A}|_{1}={\rm Tr}\,\sqrt{ { M}_{{\cal A},\bar{\cal A}}
  {M}_{\bar{\cal A},{\cal A}}}=||M_{\cal A,\bar{A}}||_1=
  \sum_{\alpha}\tilde{\sigma}_\alpha^{\cal A,\bar{A}}\,,
\label{eq:l1measure}
\end{equation}
with $\tilde{\sigma}_\alpha^{\cal A,\bar{A}}$ the singular
values of the matrix $M_{\cal A,\bar{A}}$ (in comparison,
$||M_{\cal A,\bar{A}}||_2^2
=\sum_\alpha(\tilde{\sigma}_\alpha^{\cal A,\bar{A}})^2$). If each
site in ${\cal \bar{A}}$ has at most one neighbor in ${\cal A}$
(the opposite may not hold), the rows of $M_{\cal A,\bar{A}}$ will
be orthogonal and the singular values will be
$\tilde{\sigma}_i^{\cal A,\bar{A}}=\sqrt{n_i^{\bar{A}}}$, leading
to
\begin{equation}
 |\partial {\cal A}|_{1}=\sum_{i \in {\cal A}}
 \sqrt{ ({ M}_{{\cal A},\bar{\cal A}}\,{ M}_{\bar{\cal A},{\cal A}})_{ii}}\,
=\sum_{i\in{\cal A}}\sqrt{n_i^{\cal \bar{A}}}\,,
\label{eq:l1measure2}
\end{equation}
which will differ from (\ref{eq:l2measure}) if $n_i^{\cal
\bar{A}}>1$. In general, Eq.\ (\ref{eq:l1measure2}) may provide
a rough approximation to the area (\ref{eq:l1measure}).
Interestingly, in an isotropic hypercubic lattice in $d$
dimensions, the approximation (\ref{eq:l1measure2}) is just
proportional to the euclidean area for large planar surfaces, both
parallel and tilted (with an angle of $\pi/4$ with respect to the
principal axes of the lattice), which is not true in the tilted
case neither for $|\partial{\cal A}|_2$ nor $|\partial{\cal
A}|_1$ (see next section).

In general, for two contiguous subsystems ${\cal B}$, ${\cal C}$,
previous expressions generalize to
 \begin{equation}
{\cal E}^{{\cal N}}_{{\cal B},{\cal C}}\propto |\partial {\cal B}
\cap\partial{\cal C}|_1\,,
\label{E1bc}
\end{equation}
 at leading order in $\lambda$, where
\begin{equation}
  |\partial {\cal B}\cap\partial{\cal C}|_{1}=||M_{\cal B,C}||_1=
  \sum_{\alpha}\tilde{\sigma}_\alpha^{\cal B,C}\,,
\label{eq:l1measuree}
\end{equation}
is a measure of the contacting area between ${\cal B}$ and ${\cal
C}$. Again, if each mode  in ${\cal C}$ is linked with at most one
mode in ${\cal B}$, $\tilde{\sigma}_{i}^{\cal B,C}=\sqrt{n_i^{\cal
C}}$, where $n_i^{\cal C}=(M_{\cal B,C}M_{\cal C,B})_{ii}$ is the
number of first neighbors of $i$ in ${\cal C}$.

Previous geometric-like expressions can of course be also applied
to {\it a general constant coupling}
$\Delta^-_{ij}=\frac{1}{2}\Delta^- M_{ij}$, where $M_{ij}=1$ if
pairs $i,j$ are linked by the coupling and $0$ otherwise, leading
to effective areas $|{\cal\partial A}|_1=||M_{\cal A,\bar{A}}||_1$
and $|{\cal \partial A}|_2=||M_{\cal A,\bar{A}}||_2^2$. On the
other hand, they cannot be directly applied to higher order
effects, like those depending on Eq.\ (\ref{ecFDelta2}),  as
discussed in the next section.

\section{Examples and asymptotic expressions}
We will now use the present formalism to obtain analytic
asymptotic expressions for ${\cal E}_{{\cal A},\bar{\cal A}}$ and
${\cal E}_{{\cal B},\bar{\cal C}}^{\cal N}$ for typical subsystems
${\cal A}$, ${\cal B}$ and ${\cal C}$ of a two-dimensional
lattice,  which will be compared with the exact numerical results
and the estimations (\ref{E2})--(\ref{E1bc}). We first consider
the ground state of a bosonic square lattice with attractive first
neighbor couplings
\[\Delta^\pm_{\bm{ij}}=\frac{1}{2}
\sum_{\mu=x,y}\Delta^\pm_\mu(\delta_{\bm{i},\bm{j}+\bm{u}_\mu}
+\delta_{\bm{i},\bm{j}-\bm{u}_\mu})\]
where $\bm{u}_\mu$ denotes the unit vector along the $\mu$ axis.
We have considered in Figs.\ (\ref{f1}--\ref{f6}) the isotropic case
$\Delta^\pm_{x}=\Delta^\pm_y=\Delta^\pm$, with
$\Delta^-/\Delta^+=2/3$, and a uniform single mode energy
$\lambda_i=\lambda$. Away from the critical point
$\lambda=\lambda_c$ (where, for fixed $\Delta^\pm$, the lowest
energy $\omega_\alpha$ vanishes), the system is gapped and a
finite correlation length $\xi <\infty$ is expected.
Approximately, $\lambda_c\approx 2(\Delta^++|\Delta^-|)$ (exact
result for the cyclic case \cite{MRC.10}).

\subsection{Entanglement between complementary subsystems}

\begin{figure}
  \centering
  \scalebox{.7}{\includegraphics{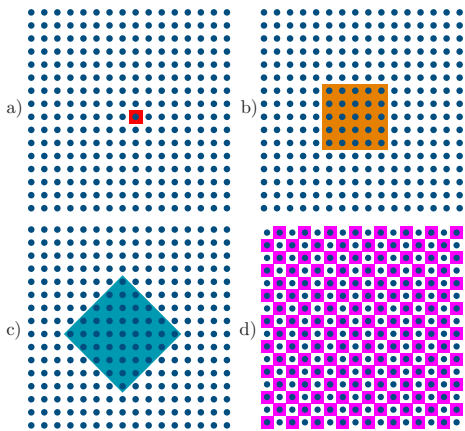}}
\caption{(Color online) The complementary partitions considered in Eqs.\
(\ref{E1234}). a) Single site, b) square block parallel to the
principal lattice axes, c) tilted square block and d)
checkerboard.}
 \label{f1}
\end{figure}

We first consider the four global $({\cal A,\bar{A}})$
bipartitions depicted in Fig. \ref{f1}. Eqs.\ (\ref{Sw}),
(\ref{ftap}) and (\ref{ecFDelta}) lead to analytic asymptotic
expressions for the corresponding singular values $\sigma^{\cal
A,\bar{A}}_\alpha$. At lowest order, their number is just the
number of sites at the border. Defining the basic single link
singular values
\[\sigma_\mu=|\Delta^-_\mu|/(4\lambda)\,,\;\;\mu=x,y,\]
in the case of a single site (Fig.\ 1a) we obtain
\begin{equation}
\sigma_{i,\bar{i}}\approx \sqrt{2(\sigma_x^2+\sigma_y^2)}\,.\label{ss}
\end{equation}
In the rectangular $n_x\times n_y$ block 1b (parallel to the
principal axes),  there are three different singular values,
corresponding to the horizontal and vertical sides and the four
corners, given below with their multiplicity:
\begin{equation}
(\sigma^{\cal A\bar{A}}_\alpha)^2\approx
\left\{\begin{array}{rcl}
\sigma_y&,& 2(n_x-2)\\
\sigma_x&,& 2(n_y-2)\\ \sqrt{\sigma_x^2+\sigma_y^2}&,& 4
\end{array}\right. \,.
\label{rb}
\end{equation}
In the  $n\times n$ square block 1c tilted $45^{\rm o}$ with
respect to the principal axes, we obtain, by means of a discrete
Fourier transform and neglecting corner effects (see Eq.\
(\ref{B2})),
\begin{equation}
\sigma^{\cal A,\bar{A}}_k \approx
 \sqrt{\sigma_x^2+\sigma_y^2+2\sigma_x\sigma_y\cos
 {\textstyle\frac{2\pi k}{m}}}\,,\;
 \label{stilt}
 \end{equation}
where $k=1,\ldots,m$ and $m=4n-4$ is the number of sites at the
border. Corner effects will affect essentially just 4 of these
eigenvalues with $O(1)$  corrections.

Finally, in the checkerboard partition 1d, an exact analytic
expression for the $n_xn_y/2$ singular values is available in the
cyclic case again by means of a discrete Fourier transform  (see
Eq.\ (41) in \cite{RCM.11}):
\begin{equation}\sigma^{\cal A,\bar{A}}_{\bm{k}}
\approx 2|\sum_{\mu=x,y}\sigma_\mu\cos {\textstyle\frac{2\pi
k_\mu}{n_\mu}|}\,,\end{equation} where $\bm{k}=(k_x,k_y)$ with
$k_x=1,\ldots,n_x$, $k_y=1,\ldots,n_y/2$.

\begin{figure}
  \centering
 \scalebox{.9}{\includegraphics{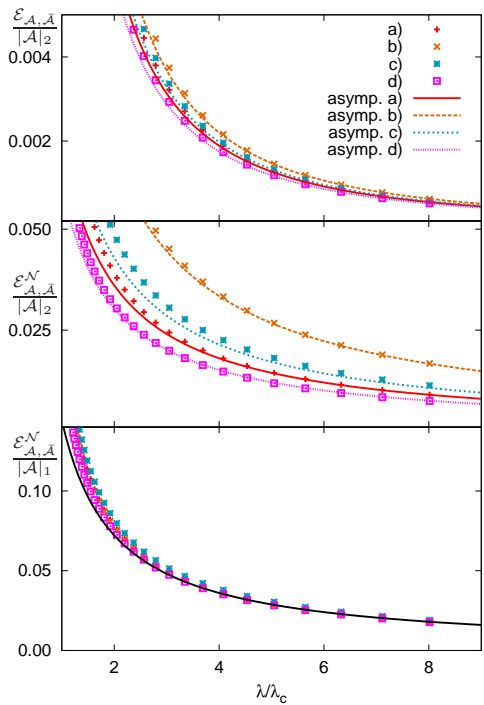}}
\caption{(Color online) Exact and asymptotic (Eqs.\ (\ref{E1234})-(\ref{N1234}))
results for the scaled entanglement entropy ${\cal E}_{\cal
A,\bar{A}}$ (top) and logarithmic negativity ${\cal E}_{\cal
A,\bar{A}}^{\cal N}$ (center and bottom) of the four bipartitions
$a,b,c,d,$ of Fig.\ \ref{f1}, as functions of the ratio
$\lambda/\lambda_c$. In the top and central panels results were
scaled with the boundary measure  $|\partial A|_2$ ($4,4n,8n,2n^2$
in $a,b,c,d$, according to Eqs.\
(\ref{eq:l2measure})--(\ref{areaf})), which is seen to provide an
adequate scaling for ${\cal E}_{\cal A,\bar{A}}$ but not ${\cal
E}_{\cal A,\bar{A}}^{\cal N}$. The latter scales accurately with
the measure $|\partial A|_1$
($2,4n,\frac{16}{\pi}n,\frac{8n^2}{\pi^2}$ according to Eqs.\
(\ref{eq:l1measure})--(\ref{areaf})), as verified in the bottom
panel.  Results correspond to a $30\times 30$ lattice with
$\Delta^-/\Delta^+=2/3$ and $10\times 10$ blocks in b) and c).}
  \label{f2}
\end{figure}

These expressions, together with Eqs.\ (\ref{Sw}), (\ref{Sp}),
(\ref{ftap}) and (\ref{Np}), allow to easily obtain the asymptotic
values of the entanglement entropy and negativity of the present
bipartitions for large $\lambda$ and $n$. For instance, in the
isotropic case $\Delta_\mu^\pm=\Delta^\pm$
considered in the figures, setting $\sigma=\sigma_\mu$ and
neglecting corner and border effects (which just add terms of
relative order $n^{-1}$), we obtain
\begin{subequations}
\begin{eqnarray}
{\cal E}^a_{i,\bar{i}}&\approx& -4\sigma^2\log_2
{\textstyle\frac{4\sigma^2}{e}}\,,\label{E1a}\\
{\cal E}^b_{\cal A,\bar{A}}&\approx& -4n\sigma^2
\log_2{\textstyle\frac{\sigma^2}{e}}\,,\label{E2b}\\
{\cal E}^c_{\cal A,\bar{A}}&\approx& -8n\sigma^2 \log_2
\sigma^2=-(4n)2\sigma^2\log_2({\textstyle\frac{e}{2}\frac{2\sigma^2}{e}})
\,,\label{E3c}\\
{\cal E}^d_{\cal A,\bar{A}}&\approx&-2n^2\log_2(\sigma^2 e)=
-(\frac{n^2}{2})4\sigma^2\log_2({\textstyle\frac{e^2}{4}\frac{4\sigma^2}{e}})\,.
\label{E4d}
\end{eqnarray}
\label{E1234}
\end{subequations}
for the entanglement entropy of the single mode, the parallel and
tilted $n\times n$ square blocks and the $n\times n$ checkerboard
of Fig.\ \ref{f1}. We have replaced sums over $k$ in $c$--$d$  by
integrals ($\sum_{k=1}^n f(\frac{2\pi k}{n})\approx
\frac{n}{2\pi}\int_0^{2\pi}f(u)du$). Note that in the chekerboard
case the entanglement entropy scales with the ``volume''  $n^2$ of
${\cal A}$ rather than the ``area'' $n$, since all links are
broken by the partition (maximally entangled bipartition
\cite{RCM.11}).

The corresponding values of the scaled logarithmic negativity
$\tilde{\cal E}^{{\cal N}}_{\cal A,\bar{A}}={\cal E}^{{\cal
N}}_{\cal A,\bar{A}}/[2\log(e)]$ are
\begin{subequations}
\begin{eqnarray}
\tilde{\cal E}^{{\cal N}\,a}_{i,\bar{i}}&\approx&  2\sigma=\sqrt{4}\sigma\,,
\label{N1}\\
\tilde{\cal E}^{{\cal N}\,b}_{\cal A,\bar{A}}&\approx& 4n\sigma\,,
\label{N2}\\
\tilde{\cal E}^{{\cal N}\,c}_{\cal A,\bar{A}}&\approx& \frac{16}{\pi}n\sigma=
(4n)\sqrt{2}{\textstyle\frac{2\sqrt{2}}{\pi}}\,\sigma\,,\label{N3}\\
\tilde{\cal E}^{{\cal N}\,d}_{\cal A,\bar{A}}&\approx& \frac{8n^2}{\pi^2}\sigma
=(\frac{n^2}{2})\sqrt{4}({\textstyle\frac{2\sqrt{2}}{\pi}})^2\,\sigma\,.
\label{N4}
\end{eqnarray}
\label{N1234}
\end{subequations}
The last expressions in (\ref{E1234})--(\ref{N1234}) indicate the
way to read them.  They are of the  form
\begin{eqnarray}
{\cal E}_{\cal A,\bar{A}}&\approx& -Lm\,\sigma^2\,
\log(\alpha^j{\textstyle\frac{m\sigma^2}{e}})\,,\label{Sapp}\\
\tilde{\cal E}^{\cal N}_{\cal A,\bar{A}}&\approx&
 L\sqrt{m}\,\beta^j\sigma\,,\label{Napp}\end{eqnarray}
where $L$ is the number of modes at the border of ${\cal A}$
($L=1,4n,4n,n^2/2$), $m$ is the number of connections with the
environment ${\cal \bar{A}}$ per mode at the border ($m=4,1,2,4$),
i.e., the number of links per mode broken by the partition, and
$\alpha^j$, $\beta^j$, with $\alpha=e/2\approx 1.36$,
$\beta=2\sqrt{2}/\pi\approx 0.9$, are {\it geometric} correction
factors for the tilted ($j=1$) and checkerboard $(j=2)$ cases
($j=0$ for the single mode and parallel square). We can easily
identify from (\ref{Sapp})--(\ref{Napp}) the boundary measures of
Eqs.\ (\ref{E2})--(\ref{E1}):
\begin{equation}
|{\cal \partial A}|_2=Lm,\;\;|{\cal \partial A}|_1=L\sqrt{m}\beta^j\,.
\label{areaf}\end{equation}

Comparison with the exact numerical results (Fig.\ \ref{f2})
indicate that all these asymptotic expressions are actually quite
accurate already for $\lambda\agt 4\lambda_c$. The scaling of
${\cal E}^{\cal N}_{\cal A,\bar{A}}$ with the area $|\partial{\cal
A}|_1$ rather than $|\partial{\cal A}|_2$ is clearly verified.
Moreover, this scaling is more accurate than that of the
entanglement entropy ${\cal E}_{\cal A,\bar{A}}$ with
$|\partial{\cal A}|_2$, since the latter  contains  in $1c-1d$ an
additional geometric correction $Lm\log(\alpha^j m)\sigma^2$ (Eq.\
(\ref{Sapp})), not comprised in (\ref{E2}). Note also that in the
case of the tilted block, $|{\cal \partial A}|_2=2L$ and $|{\cal
\partial A}|_1=L\sqrt{2}\beta=(4/\pi)L$ are, respectively, larger
and smaller ($90\%$) than the geometric perimeter $\sqrt{2}L$.

We may also rapidly determine with Eqs.\ (\ref{rb}) and
(\ref{Sapp})--(\ref{Napp}) the corner effects in case 1b. The
actual asymptotic expressions for the finite $n\times n$ parallel block are
\begin{eqnarray}
 {\cal E}^b_{\cal A,\bar{A}}&\approx&
-4(n-1)\sigma^2\log_2{\textstyle\frac{\sigma^2}{e}-4\sigma^2\log_2
\frac{4\sigma^2}{e}}\,,\label{E22}\\
\tilde{\cal E}^{{\cal N}\,b}_{\cal A,\bar{A}}&\approx&
4(n-1)\sigma+4(\sqrt{2}-1)\sigma\,,
\label{N22}
\end{eqnarray}
where the first term is proportional to the number of sites at the
border, $4(n-1)$, and the second represents the positive
correction arising from the four corners, reflecting their increased
coupling with the environment $\bar{\cal A}$.

\subsection{Non-complementary subsystems}
\begin{figure}
  \centering
    \scalebox{.45}{\includegraphics{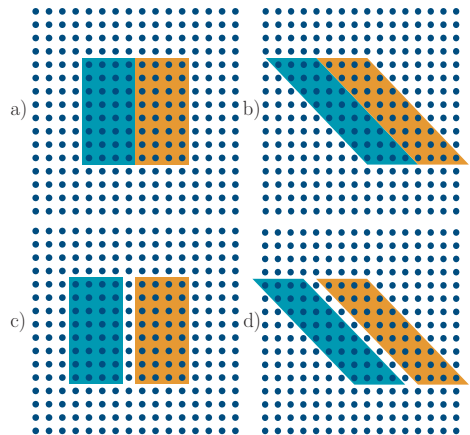}}
\caption{(Color online) Non complementary subsystems: Contiguous (top) and
one-site separated (bottom) blocks, with contacting sides parallel
(left) and tilted (right) with respect to the principal axes. The
negativity and its size dependence are determined by both
separation and slope of the contacting boundary.}
  \label{f3}
\end{figure}

Let us now consider the non-complementary subsystems of Fig.\
\ref{f3}. For contiguous parallel blocks  contacting at $n$ sites,
$F^-_{\cal B,C}$ has $n$ identical singular values
\begin{equation}
\sigma^{\cal B,C}_\alpha=\sigma_x\,.
\end{equation}
In the case of contiguous blocks with contacting surfaces tilted
$45^{\rm o}$ with respect to the principal axes, we obtain,
neglecting edge effects, the same expression (\ref{stilt}) for the
$\sigma_k^{\cal B,C}$, with $m$ replaced by the number of
contacting sites $n$ and $k=1,\ldots,n$. In the isotropic case we
then obtain, using Eqs.\ (\ref{ftap})--(\ref{Np}),
\begin{eqnarray}
\tilde{\cal E}^{{\cal N} a}_{\cal B,\bar{C}}&\approx&
n\,\sigma\,,\label{NBCappr}\\
\tilde{\cal E}^{{\cal N} b}_{\cal B,\bar{C}}&\approx&
\frac{4}{\pi}n\sigma=n\sqrt{2}{\textstyle\frac{2\sqrt{2}}{\pi}}
 \,\sigma\,,\label{NBCappt}\end{eqnarray}
for the logarithmic negativity of parallel and tilted contiguous
blocks, which are clearly of the form (\ref{Napp}) or (\ref{E1}):
$|{\cal\partial {\cal B}\cap \partial{\cal C}}|_1=n$ and $4n/\pi$
respectively. Tilted boundary surfaces exhibit a larger
entanglement per contacting site due to the increased
connectivity.

\begin{figure}
 \centering \scalebox{.9}{\includegraphics{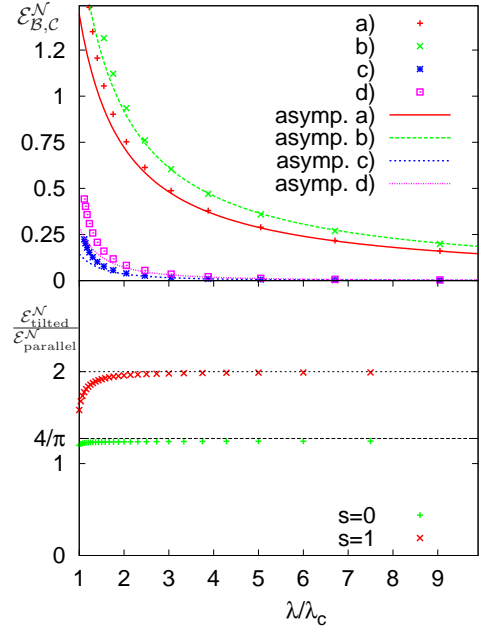}}
\caption{(Color online) Top: Exact and asymptotic logarithmic negativities (Eqs.\
(\ref{NBCappr})--(\ref{NBCappt})) for subsystems of the type of fig.\
\ref{f3} (for $10\times 10$ blocks) as functions of
$\lambda/\lambda_c$. Tilted blocks exhibit a larger negativity per
contacting site. Bottom: The tilted to parallel logarithmic
negativity ratio for separations $s=0$ (a--b) and $1$ (c--d). 
It is asymptotically $4/\pi$ in the contiguous
case and $2$ for one site separation.}
  \label{f4}
\end{figure}

\begin{figure}
  \centering
\scalebox{.9}{
\includegraphics{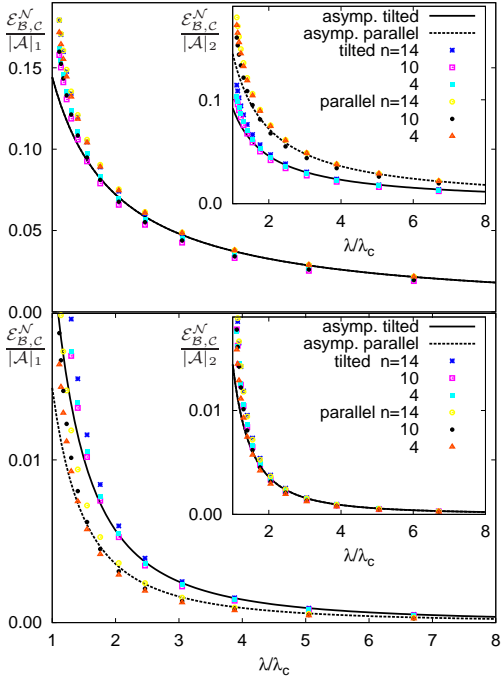}}
\caption{(Color online) Top: Asymptotic and exact values of the logarithmic
negativity of two contiguous $n\times n$ blocks with parallel
and tilted boundary surfaces,
scaled with $|{\cal \partial A}|_1$ in the main panel and $|{\cal
\partial A}|_2$ in the inset, with ${\partial\cal A}=
{\partial\cal B}\cap{\partial\cal C}$. Bottom: Same details
for two blocks separated by one site. The appropriate scaling is
verified to be $|{\cal \partial A}|_1$ in the top panel and $|{\cal \partial
A}|_2$ in the bottom panel.}
  \label{f5}
\end{figure}

In the case of blocks separated by one site, we should use instead
the full Eq.\ (\ref{Gw}) with the {\it second order} expression
(\ref{ecFDelta2}). For parallel blocks with $n$ sites at
separation $s=1$, the $n$ negative eigenvalues of the matrix
(\ref{eqneg0})  become, neglecting edge effects and setting
$\sigma^+_\mu=|\Delta^+_\mu|/(4\lambda)$,
\begin{equation}\tilde{f}_\alpha^{\cal B,C}\approx
-(2\sigma^+_x\sigma_x-\sigma_x^2)\,.\label{Nel1}
\end{equation}
For blocks separated by one site through a $45^{\rm o}$ tilted
surface of $n$ modes, a discrete Fourier transform leads,
neglecting edge effects, to  (see Eq.\ (\ref{B2}))
\begin{eqnarray}
\tilde{f}_k^{\cal B,C}&\approx&-\{2[
\alpha_{xy}^2+\alpha_x^2+\alpha_y^2
+2\alpha_{xy}(\alpha_x+\alpha_y)
\cos{\textstyle\frac{2\pi k}{n}}\nonumber
\\&&+2\alpha_x\alpha_y\cos{\textstyle\frac{4\pi k}{n}}]^{1/2}-\sigma_k^2\}\,,
\label{Nel2}
\end{eqnarray}
where $\alpha_\mu=\sigma^+_\mu\sigma_\mu$,
$\alpha_{x,y}=\sigma_x^+\sigma_y+\sigma_y^+\sigma_x$ and
$\sigma_k$ denotes the expression (\ref{stilt}) for $m=n$. In the
isotropic case, Eq.\ (\ref{Nel2}) becomes just
$4\sigma(2\sigma^+-\sigma)\cos^2\frac{\pi k}{n}$. For the parallel
and tilted subsystems $c$--$d$ of Fig.\ \ref{f3} we then obtain,
replacing sums by integrals and assuming $\sigma\leq 2\sigma^+$,
\begin{subequations}
\label{EC}
 \begin{eqnarray}
\tilde{\cal E}^{{\cal N} c }_{\cal B,\bar{C}}&\approx&
n\sigma(2\sigma^+-\sigma)\,,\label{EC1}\\
\tilde{\cal E}^{{\cal N} d }_{\cal B,\bar{C}}&\approx&
2n\sigma(2\sigma^+-\sigma)\,.\label{EC2}
\end{eqnarray}
\end{subequations}
Hence, the logarithmic negativity of the tilted case is,
remarkably,  {\it twice} that of parallel blocks when separated by
one site, instead of $4/\pi\approx 1.27$ as in the contiguous case
(Fig.\ \ref{f4}). Since they are a second order effect, Eqs.\
(\ref{EC}) are not of the form (\ref{Napp}) but rather
\begin{eqnarray}
\tilde{\cal E}^{\cal N }_{\cal B,\bar{C}}&\approx&
 Lm\sigma(2\sigma^+-\sigma)\,, \,\end{eqnarray}
if $m$ is again the number of connections with the environment per
mode. They scale, therefore, with the measure $|{\cal \partial
A}|_2$ (Fig.\ \ref{f5}). For larger
separations $s$ the negativity vanishes at second order in
$\Delta/\lambda$, as $F^-_{\cal B,C}$ will be of higher order
while the counter terms $G_{\cal B}$ and $G_{\cal C}$ remain
of second order for sites at the surface. Consequently, the
negativity becomes vanishingly small for $s\geq 2$.

\begin{figure}
  \centering
 \scalebox{.6}{\includegraphics{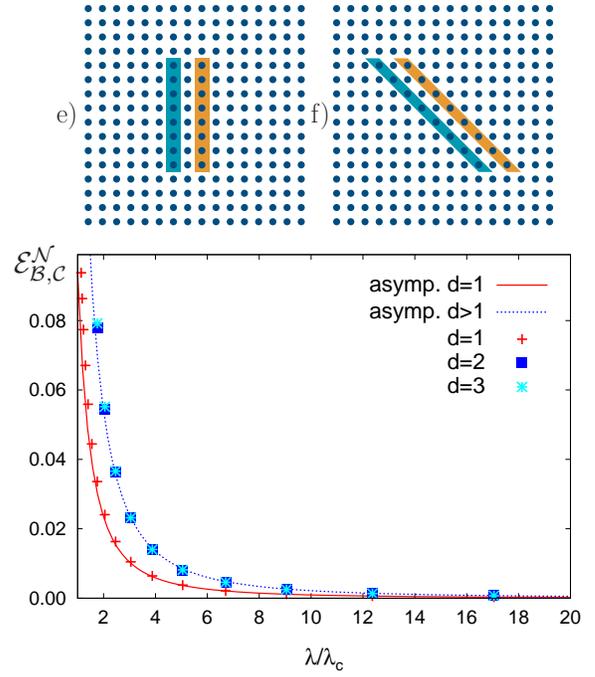}}
\caption{(Color online) Top: Parallel and tilted lines separated by one site. Due
to the extra interaction with the environment, the associated
negativity (Eqs.\ (\ref{ED})) is lower than that of the
corresponding blocks $(c,d)$ of Fig.\ \ref{f3} (Eqs.\ (\ref{EC})),
as appreciated in the bottom panels for the parallel case.}
  \label{f6}
\end{figure}

We finally remark that previous expressions are independent (at
leading order) of the width $d$ of the blocks (assumed finite),
{\it provided} $d\geq 2$. In the case of two lines ($d=1$, Fig.\
\ref{f6}), the extra interaction with the environment at the other
side of the line leads to an additional negative second order
contribution in Eq.\ (\ref{Gex}). Hence, while it can be neglected
in the case of contiguous lines, it will double the negative term
in Eqs.\ (\ref{EC}) in the case of lines separated by one site,
leading to the lower values
\begin{subequations}
\label{ED}
 \begin{eqnarray}
\tilde{\cal E}^{{\cal N} e }_{\cal B,\bar{C}}&\approx&
n\sigma(2\sigma^+-2\sigma)\,,\label{ED1}\\
\tilde{\cal E}^{{\cal N} f }_{\cal B,\bar{C}}&\approx&
2n\sigma(2\sigma^+-2\sigma)\,,\label{ED2}
\end{eqnarray}
\end{subequations}
which are now valid for $\sigma\leq \sigma^+$. In this case
negativity will vanish at second order if $\sigma>\sigma^+$.
Edge effects in Eqs.\ (\ref{EC})--(\ref{ED})  are also of
second order and lead, using Eq.\ (\ref{Gw}), to a negative
correction $-2\sigma^2$.

All present expressions can
be directly extended to three dimensions if present
subsystems are extended parallel-wise along the $z$ axis,
replacing $n$ by $n n_z$.

\subsection{The fully connected case}

The evaluation of singular values is also straightforward in the
opposite case of a fully and uniformly connected system of $n$ modes 
\cite{BDV.06,WVB.10,MRC.082,MRC.10} (LMG type model \cite{RS.80}), where
 \begin{equation}
\Delta^\pm_{\bm{ij}}=(1-\delta_{\bm{ij}}\frac{\Delta^{\pm}}{n-1})\,.
 \label{deltL}\end{equation}
Here we can also compare with full exact results, since it is
exactly and analytically solvable \cite{WVB.10,MRC.10}. The
present system can be used to describe entanglement between
systems whose separation is small in comparison with the
correlation length.

In the present case, the matrices $F^\pm_{ij}$  are obviously constant
for $i\neq j$, i.e.,
\begin{equation}
F^\pm_{\bm{ij}}=F^{\pm}_0\delta_{\bm{ij}}+F^\pm_1\,,
\end{equation}
and the entanglement between disjoint subsystems ${\cal B}$ and
${\cal C}$ will just depend on the number of sites in ${\cal B}$
and in ${\cal C}$, being independent of their separation or shape.

The matrix $F^-_{\cal B,C}$ will then have just a
{\it single} non-zero singular value $\forall$ disjoint ${\cal B, C}$,
namely (see Appendix)
\begin{equation}
\sigma^{\cal B,\cal C}=\sqrt{n_{\cal B} n_{\cal
 C}}|F^-_1|\label{LS}\,.\end{equation}
In the approximation (\ref{Sw}) we then  obtain a {\it single}
non-zero symplectic eigenvalue for {\it any} global bipartition ${\cal
A,\bar{A}}$,
\begin{equation}
f^{\cal A}\approx n_{\cal A}n_{\cal \bar{A}}(F_1^-)^2\label{Lf}\,,
\end{equation}
where $n_{\cal \bar{A}}=n-n_{\cal A}$, leading to
\[{\cal E}_{\cal A,\bar{A}}
\approx -n_{\cal A}n_{\cal \bar{A}} |F_1^-|^2\log(n_{\cal A}
 n_{\cal \bar{A}} (F_1^-)^2/e)\,,\]
which corresponds to an area $|{\cal \partial A}|_2=n_{\cal A}n_{\cal
\bar{A}}$ in (\ref{E2}) (here $M_{ij}=1$ for $i\neq j$).

Similarly, we obtain a {\it single} negative symplectic eigenvalue
for any pair of subsystems ${\cal B,C}$, given by (\ref{LS}) or,
in the complete approximation (\ref{eqneg0})--(\ref{Gw}), by
\begin{equation}
\tilde{f}^{\cal B,C}\approx -\sqrt{n_{\cal B}n_{\cal {C}}}
|F_1^-|+{\textstyle\frac{1}{2}|F^-_1|^2(n_{\cal B}
n_{\cal \bar{B}}+n_{\cal C}n_{\cal \bar{C}})}\label{Lfs}\,,
\end{equation}
with ${\cal E}^{\cal N}_{\cal B,C}=-\tilde{f}_{\cal B,C}$. The
second term in (\ref{Lfs}) becomes important for small subsystems
in a large environment ($n_{\cal B},n_{\cal C}\ll n$). Otherwise
it can be neglected, in which case (\ref{Lfs}) corresponds to
$|{\cal \partial A}|_1=\sqrt{n_{\cal B}n_{\cal C}}$ in
(\ref{E1})--(\ref{E1bc}). For the scaling (\ref{deltL}), $F^-_1$
is proportional to $n^{-1}$, so that Eqs.\ (\ref{LS})--(\ref{Lfs})
remain finite for large $n$. The scaling is then again as in Eqs.\
(\ref{Sapp})--(\ref{Napp}) with $L=1$, $j=0$ and $m=n_{\cal
A}n_{\cal \bar{A}}$ for global partitions or $m=n_{\cal B}n_{\cal
C}$ for a pair of subsystems.

The previous picture is, remarkably, also that of the exact
treatment, where  there is a {\it single} non-zero symplectic
eigenvalue $f^{\cal A}$ for any subsystem ${\cal A}$, given  by
\begin{equation}
f^{\cal A}={\textstyle\sqrt{\frac{1}{4}
+F^+_1 n_{\cal A}n_{\cal \bar{A}}/n}-\frac{1}{2}}\,,\label{fex}
\end{equation}
(see ref.\ \cite{MRC.10} and Appendix \ref{C}). Here we have used the local
basis where $F^{\pm}_0=0$ in Eq.\ (\ref{deltL}), in which case Eq.\
(\ref{Fex}) leads to
\[{F^+_1}^2+F^+_1/n=(F^-_1)^2\,.\]
A first order expansion of (\ref{fex}) in $F^+_1$ leads then to
$f^{\cal A}\approx F^+_1 n_{\cal A}n_{\cal \bar{A}}/n$, which
coincides with Eq.\ (\ref{Lf}) since for weak coupling
$F^+_1/n\approx (F^-_1)^2$.

In the same way, the exact partial transpose of ${\cal D}_{BC}$
has a {\it single} negative symplectic eigenvalue \cite{MRC.10}
\begin{equation}
\tilde{f}^{\cal B,C}=
{\textstyle\sqrt{\frac{1}{4}+\gamma_{\cal B,C}F_1^+-
\sqrt{F_1^+(\beta_{\cal B,C}+\gamma_{\cal B,C}^2F_1^+)}}-\frac{1}{2}}\,.
\label{fexx}
\end{equation}
where $\beta_{\cal B,C}=n_{\cal B}n_{\cal C}/n$, $\gamma_{\cal
B,C}=\frac{1}{2}(n_{\cal B}+n_{\cal C})(n-n_{\cal B}-n_{\cal
C})/n+2\beta_{\cal B,C}$ (see appendix \ref{C}). Expansion of
(\ref{fexx}) up to first order in $F_1^+$ leads then exactly to
Eq.\ (\ref{Lfs}) setting $F_1^+\approx n(F_1^-)^2$. The present
approximate scheme allows then to immediately determine the weak
coupling expressions (\ref{Lf})--(\ref{Lfs}) and to  rapidly
identify their behavior with sizes $n_{\cal A}$, $n_{\cal B}$, $n_{\cal C}$. The
exact value of $F_1^\pm$ (to be inserted in
(\ref{fex})--(\ref{fexx})) is given in the appendix (Eq.\
(\ref{fapl})).  Up to first order in $\Delta_-$ we obtain
$F_1^-\approx \frac{\Delta^-}{2(n-1)\lambda}$.

\section{Conclusions}

We have shown how entanglement properties of weakly correlated
gaussian states can be recast in terms of the singular values of a
sub-block of the generalized contraction matrix associated with
the state. This allows to obtain in a quite simple way analytic
expressions for both the entanglement entropy between
complementary subsystems and the logarithmic negativity for non
complementary subsystems, which imply distinct area laws for these
two quantities in the case of short range or constant couplings.
Several illustrative examples were considered, which show the
dependence of these laws on the geometry, connectivity and
separation between the subsystems. A final comment is that through
application of the bosonic RPA formalism
\cite{MRC.10,RCM.11,MRC.08} or other bosonization treatments,
\cite{WVB.10,RS.80}, the present scheme can be applied to
weakly interacting spin systems. Moreover, it can in principle be
also implemented in phases exhibiting symmetry-breaking at the
mean field level (i.e., fields below the critical field in
attractive $XY$ or $XYZ$ chains) away from the critical field,
provided the proper multiplicity corrections accounting for the
different degenerate mean fields \cite{MRC.10} is taken into
account. Such application is currently being investigated. 

The authors acknowledge support of CONICET (NC,JMM) and CIC (RR) of
Argentina.

\appendix

\section{Perturbative expansions for the symplectic eigenvalue problem}
\label{app:perturb}
In this work we have used perturbative results which are not necessarily 
trivial and which can be obtained following 
techniques similar to those employed in the perturbative
diagonalization of the Dirac equation. We start with the
symplectic diagonalization of the contraction matrix ${\cal
D}_{\cal A}$ of a subsystem ${\cal A}$, which leads to the system
\begin{eqnarray}
  F^{+}_{\cal A}U_f-  F^{-}_{\cal A}\bar{V}_f&=& f \,U_f\,,\label{B1a}\\
 \bar{F}^{-}_{\cal A}U_f  -(\bm{1}+\bar{F}^+_{\cal A})\,\bar{V}_f&=& f \,\bar{V}_f
 \label{B1b}\end{eqnarray}
where $(^{U_f}_{\bar{V}_f})$ is the symplectic eigenvector associated with the
eigenvalue $f$. Eq.\ (\ref{B1b}) allows to write $V_f$ as
\begin{equation}\bar{V}_f=[\bm{1}(1+f)+\bar{F}^+_{\cal A}]^{-1}\bar{F}^-_{\cal A}
 U_f\,.\label{B2a}\end{equation}
Replacing (\ref{B2a}) in (\ref{B1a}) leads to the equivalent non-linear reduced
diagonalization problem
\[\{F^+_{\cal A}-F^-_{\cal A}
[\bm{1}(1+f)+\bar{F}^+_{\cal A}]^{-1}\bar{F}^-_{\cal A}\}U_f=f
 U_f\,.\]
For small $F^{\pm}$, in agreement with the hypothesis that the
state is weakly correlated, the symplectic eigenvalues $f$ are
small. Hence, at leading order $\bar{V}_f \approx\bar{F}^{-}_{\cal
A}U_f$ and  we obtain the reduced standard eigenvalue equation
\begin{equation}
(F^{+}_{\cal A}-{F}^{-}_{\cal A}\bar{F}^{-}_{\cal A})U_f=f U_f\,,\label{eqc1}
\end{equation}
which leads to Eq.\ (\ref{eq:FplusWCS}) and implies Eq.\
(\ref{eqneg0}). If ${\cal A}$ is the whole system and the latter
is assumed to be in the ground state of $H$, all $f$ vanish and
the relation $F^+\approx F^- \bar{F}^-$ (Eq.\ (\ref{F1})) is
obtained.

Let us now consider the Hamiltonian (\ref{eq:H1}).  The symplectic
diagonalization of ${\cal H}$ entails the standard diagonalization
of ${\cal M}{\cal H}$ and  leads to the system
\begin{eqnarray}
 (\Lambda-\Delta^{+})U_\omega- \Delta^{-}\bar{V}_\omega&=& \omega \,U_\omega\\
 \bar{\Delta}^{-}U_\omega  - (\Lambda-\bar{\Delta}^{+})\,\bar{V}_\omega&=&
 \omega \,\bar{V}_\omega\,.
\end{eqnarray}
In this case, $\|\Delta^-\|_{\infty}$ is considered small. For a
positive eigenvalue, the zero order approximation is obtained by
neglecting all terms proportional to $\Delta^{-}$ and
$\bar{V}_{\omega}$, which are assumed small in comparison with
$U_\omega$, $\omega$ and  $\Lambda-\Delta^+$. It leads to
$(\Lambda-\Delta^{+})U_\omega= \omega \,U_\omega$, which is a
standard hermitian eigenvalue equation for $U_\omega$. We then
obtain
\begin{eqnarray}
\bar{V}_{\omega}&=
&(\Lambda-\bar{\Delta}^++\omega\bm{1})^{-1}\bar{\Delta}^-U_\omega
\label{A6}
\\ &\approx&
{U}^{*}(\Omega+\omega\bm{1})^{-1}{U}^{t}\bar{\Delta}^-U_{\omega}\,\,,
\end{eqnarray}
(Eq.\ (\ref{eq:Vpert1})), where we have written
$\Lambda-\Delta^{+}\approx U\Omega U^\dagger$, with $\Omega={\rm
diag}(\omega_\alpha)$ the diagonal matrix of eigenvalues. It
should be noticed that if $\Lambda$ is degenerate, $\Delta^+$ will
affect $U$ considerably even if small. Ground state entanglement
will remain however small since it depends on $V$. It can be also
easily seen that expansion of Eq.\ (\ref{A6}) up to second order
in $\Delta^\pm$ leads  to Eq.\ (\ref{ecFDelta2}).

\section{\label{S} Singular values}

The {\it singular values} $\sigma_\alpha$ of an arbitrary $m\times
n$ matrix $A$ are the square root of the non-zero eigenvalues of
$AA^\dagger$ or equivalently $A^\dagger A$, which are both
positive matrices with the same non-zero eigenvalues. The singular
value decomposition implies the existence of unitary matrices $U$,
$V$ such that $A=UDV^\dagger$, with $D$ a diagonal matrix with
diagonal elements $\sigma_\alpha$ or $0$, and $U$, $V$ unitary
eigenvector matrices of $AA^\dagger$ and $A^\dagger A$:
$AA^\dagger U=UDD^\dagger$, $A^\dagger A V=VD^\dagger D$, i.e.,
$AA^\dagger U_\alpha=\sigma^2_\alpha U_\alpha$, $A^\dagger A
V_\alpha=\sigma^2_\alpha V_\alpha$ for the non-zero eigenvalues
$\sigma_\alpha$, with $V_\alpha=A^\dagger U_\alpha/\sigma_\alpha$. 
 For an hermitian $A$, $\sigma_\alpha=|\lambda_\alpha|$, with $\lambda_\alpha$ the
(non-zero) eigenvalues of $A$.

The singular values determine the matrix $m$ norm of $A$ used in
this work, defined as
\begin{equation}||A||_m=[{\rm Tr}\,(A^\dagger A)^{m/2}]^{1/m}=
 (\sum_\alpha \sigma_\alpha^m)^{1/m} \,.\end{equation}
$||A||_1$ is the trace norm, $||A||_2$  the standard
Hilbert-Schmidt norm and $||A||_\infty$ the spectral norm, which
is just the largest singular value.

The singular values $\sigma_\alpha$ of $A$ also determine the
non-zero eigenvalues of the hermitian  $(m+n)\times (m+n)$ matrix
\begin{equation}
B=\left(\begin{array}{cc}0&A\\A^\dagger &0\end{array}\right)\,,
\end{equation}
which are $\pm \sigma_\alpha$, since
$B^2=(^{AA^\dagger\;0}_{0\;\;\;A^\dagger A})$ has eigenvalues
$\sigma_\alpha^2$. Eigenvalues $\pm\sigma_\alpha$ correspond to
normalized eigenvectors $(^{\;\;U_\alpha}_{\pm
V_\alpha})/\sqrt{2}$, with $AA^\dagger U_\alpha=\sigma^2_\alpha
U_\alpha$, $V_\alpha=A^\dagger U_\alpha/\sigma_\alpha$ and
$U_\alpha^\dagger U_\beta=\delta_{\alpha\beta}$, $V_\alpha^\dagger
V_\beta=\delta_{\alpha\beta}$.

These results first imply that the non-zero eigenvalues of the
matrix (\ref{eq:FplusWCS}) are the square of the singular values
$\sigma_\alpha^{\cal A,\bar{A}}$ of $F^-_{\cal A,\bar{A}}$, as
$\bar{F}^-_{\cal \bar{A},A}=(F^-_{\cal A,\bar{A}})^\dagger$,
implying $\sigma_\alpha^{\cal A,\bar{A}}=\sigma_\alpha^{\cal
\bar{A},A}$. They also entail that the negative eigenvalues of the
matrix (\ref{eqneg0}) are minus the singular values
$\sigma_\alpha^{\cal B,C}=\sigma_\alpha^{\cal C,B}$ of $F^-_{\cal
B,C}$, when $\bar{G}_{\cal B}$ and $G_{\cal C}$ are neglected.

\section{Evaluation of singular values\label{C}}

In the {\it first order} approximation (\ref{ecFDelta}),  the
matrix $F^-_{\cal B,C}$ for first neighbor couplings and disjoint
contiguous blocks ${\cal B}$, ${\cal C}$ with $n$ contacting
sites, has elements of the form
\begin{equation}(F^-_{\cal B,C})_{ij}=f(j-i)\label{Aij}\end{equation}
if the sites are adequately ordered, where
$f(l)=\delta_{l0}\sigma_{\mu\perp}$ for parallel and
$f(l)=\sigma_x\delta_{l0}+\sigma_y\delta_{l1}$
for tilted blocks.

For blocks separated by one site, we should use the {\it second
order} approximation (\ref{ecFDelta2}), which  leads again to a
matrix of the form (\ref{Aij}), with
$f(l)=2\delta_{l0}\sigma^+\sigma$ for parallel blocks and
$f(l)=2[\sigma^+_x\sigma_x\delta_{l0}+(\sigma_x^+
\sigma_y+\sigma_y^+\sigma_x)\delta_{l1}
+\sigma^+_y\sigma_y\delta_{l2}]$ for tilted blocks. In all
previous cases, $F^-_{\cal B,C}\bar{F}^-_{\cal C,B}$ is an
hermitian matrix with elements of the form
\begin{equation}(F^-_{\cal B,C}
\bar{F}^-_{\cal C,B})_{ij}=\sum_k f(k-i)\bar{f}(k-j)=g(i-j)\,,
\label{C2}\end{equation}
if edge effects are neglected, where $g(l)=\sum_k f(k)\bar{f}(k+l)=\bar{g}(-l)$. 
Such matrix can then be exactly diagonalized
(neglecting edge effects) by a discrete Fourier
transform \cite{MRC.082}, leading to eigenvalues
$\sigma^2_k=\sum_l g(l)e^{i2\pi kl/n}$, where 
$k=0,\ldots,n-1$ and $n$ is its dimension (this result is exact if  
$g(-l)=g(n-l)$). For real $g(l)$, as in the previous cases, we then obtain 
\begin{equation}
\sigma^2_k=g(0)+2\sum_{l>0} g(l)\cos{\textstyle\frac{2\pi k}{n}}\label{B2}\,,
\end{equation}
which leads to Eqs.\
(\ref{stilt})--(\ref{Nel2}) (in the case of ${\cal C}=\bar{\cal
A}$ with ${\cal A}$ the tilted block, the final matrix $F^-_{\cal
A,\bar{A}}\bar{F}^-_{\cal \bar{A},A}$ is again of the form
(\ref{C2})).

In the fully connected case, the exact singular values (\ref{LS})
arise immediately as the matrix $F^-_{\cal B,C}$ is just a rank 1
constant matrix,  i.e., $F^-_{\cal B,C}=c$ $\forall$ $i,j$, which
therefore has a unique non-zero singular value
$\sigma=\sqrt{n_{\cal B}n_{\cal C}}|c|$: $F^-_{\cal
B,C}\bar{F}^-_{\cal C,B}$ is a  $n_{\cal B}\times n_{\cal B}$ rank
1 matrix with constant elements $n_{\cal C}|c|^2$, whose unique
non-zero eigenvalue is $n_{\cal B}n_{\cal C}|c|^2$ due to trace
conservation.

The full exact symplectic diagonalization can also be performed
(see appendix in \cite{MRC.10} for details). We quote here that the exact
symplectic eigenvalues of the reduced state of $L$ sites 
for the couplings (\ref{deltL}) are
$\sigma_1=\sqrt{(F_0^++LF_1^++
\frac{1}{2})^2-(F_0^-+LF_1^-)^2}-\frac{1}{2}$
and $\sigma_0=\sqrt{(F_0^++\frac{1}{2})^2-(F_0^-)^2}-\frac{1}{2}$
($L-1$ fold degenerate). For a pure global state, $\sigma_0=0$. In
the local basis where $F_0^-=0$, this implies $F_0^+=0$, which
leads to Eq.\ (\ref{fex}). In the same way, we obtain Eq.\
(\ref{fexx}). The exact value of
the present $F_1^+$ was also evaluated in \cite{MRC.10} in terms
of a parameter $\Delta$ ($F_1^+=\Delta/(2n)$):
\begin{equation}
F_1^+={\textstyle\frac{n(\lambda^2-\bar{\omega}^2)}
{4(n-1)\omega_0\omega_1}}\,,\label{fapl}
\end{equation}
where $\bar{\omega}=\frac{\omega_0+(n-1)\omega_1}{n}$,
$\omega_0=\sqrt{(\lambda-\Delta_x)(\lambda-\Delta_y)}$ and
$\omega_1=\sqrt{(\lambda+\frac{\Delta_x}{n-1})(\lambda+
\frac{\Delta_y}{n-1})}$, 
with $\Delta_{\pm}=(\Delta_x\pm\Delta_y)/2$.


\begin{thebibliography}{99}
\bibitem{NC.00} M.~A.\ Nielsen, I.~L.\ Chuang,
  {\it Quantum Computation and Quantum Information}
  (Cambridge Univ.\ Press, Cambridge UK, 2000).
\bibitem{BB.93} C.H. Bennett et al,
Phys.\ Rev.\ Lett.\ {\bf 70}, 1895 (1993). 
\bibitem{JL.03} R. Josza and N. Linden, Proc. R. Soc. A 459, 2011 (2003);
 G. Vidal, Phys. Rev. Lett. 91, 147902 (2003).
\bibitem{RB.01} R. Raussendorf and H.J. Briegel, Phys. Rev. Lett. 86, 5188 (2001);
 R.\ Raussendorf, D.E.\ Browne and H.J.\ Briegel,
 Phys. Rev. A {\bf 68}, 022312 (2003).
\bibitem{WP.12} C.\ Weedbrook et al, 
Rev.\ Mod.\ Phys.\ {\bf 84}, 621 (2012).
\bibitem{OA.02}  A.\ Osterloh, L.\ Amico, G.\ Falci, and R.\ Fazio,
Nature (London) 416, 608 (2002).
\bibitem{ON.02} T.J.\ Osborne, M.A.\ Nielsen, Phys.\ Rev.\ A {\bf 66},
 032110 (2002).
\bibitem{VLRK.03} G.\ Vidal, J.I.\ Latorre, E.\ Rico, and A.\ Kitaev,
   Phys. Rev. Lett.\ {\bf 90}, 227902 (2003).
\bibitem{AFOV.08} L. Amico, R. Fazio, A. Osterloh,
and V. Vedral, Rev. Mod. Phys. {\bf 80}, 517 (2008).
\bibitem{ECP.10}  J. Eisert, M. Cramer,
and M. B. Plenio, Rev. Mod. Phys. {\bf  82}, 277 (2010).
\bibitem{BBPS.96}C.H.\ Bennett,
H.J.\ Bernstein, S.\ Popescu, and B.\ Schumacher, Phys.\ Rev.\ A {\bf 53},
2046 (1996).
\bibitem{BVSW.96}C.H.\ Bennett, D.P.\ DiVincenzo, J.A.\ Smolin, W.K.\
Wootters, Phys. Rev. A {\bf 54}, 3824 (1996).
\bibitem{RC.03}P.Rungta, C.M.\ Caves, Phys.\ Rev.\ A {\bf 67} 012307 (2003);
  P.\ Rungta et al, Phys.\ Rev.\ A {\bf 64} 042315 (2001).
\bibitem{VW.02} G.\ Vidal, R. F.\ Werner, Phys.
 Rev. A {\bf 65}, 032314 (2002).
\bibitem{Pl.05}M.B.\ Plenio, Phys.\ Rev.\ Lett.\ {\bf 95}, 090503 (2005).
\bibitem{HSH.1999}
A.~S.\ Holevo, M.\ Sohma, and O.\ Hirota,
  Phys.\ Rev.\ A {\bf 59}, 1820 (1999).
\bibitem{RS.00} R.\ Simon, Phys.\ Rev.\ Lett.\ {\bf 84}, 2726 (2000).
\bibitem{AEPW.02} K.\ Audenaert, J.\ Eisert, M. B.\ Plenio,
and R. F.\ Werner, Phys.
 Rev. A {\bf 66}, 042327 (2002).
\bibitem{CEPD.06}
M. B.\ Plenio, J.\ Eisert,
 J.\ Drei\ss ig, and M.\ Cramer, Phys.\ Rev.\
Lett.\ {\bf 94}, 060503 (2005); 
M.\ Cramer, J.\ Eisert, M.~B.\ Plenio, and  J.\ Drei\ss ig,
  Phys.\ Rev.\ A {\bf 73}, 012309 (2006).
\bibitem{CEP.07}  M.\ Cramer, J.\ Eisert, and  M.~B.\ Plenio,
  Phys.\ Rev.\ Lett.\ {\bf 98}, 220603 (2007).
\bibitem{ASI.04} G.~Adesso, A.~Serafini, and F.~Illuminati,
Phys.~Rev.~Lett.~ {\bf 93}, 220504 (2004); Phys.~Rev.~A {\bf 70}, 022318 (2004);
A.~Serafini, G.~Adesso, and F.~Illuminati,
Phys.~Rev.~A {\bf 71}, 032349 (2005).
\bibitem{AI.08}
G.~Adesso,  F.~Illuminati, Phys.~Rev.~A {\bf 78}, 042310 (2008).
\bibitem{RS.80}
P.\ Ring and P.\ Schuck, {\it The Nuclear Many-Body Problem}
(Springer, NY, 1980).
\bibitem{MRC.10}
J.~M.\ Matera, R.\ Rossignoli, and N.\ Canosa,
  Phys.\ Rev.\ A  {\bf 82} 052332 (2010).
\bibitem{MRPR.09}
S.\ Marcovitch, A.\ Retzker, M.~B.\ Plenio,  and B.\ Reznik,
  Phys. Rev. A {\bf 80}, 012325 (2009).
\bibitem{CC.04} P.\ Calabrese and J.\ Cardy,
J.\ Stat.\ Mech.:\ Theory Exp. (2004), P06002.
\bibitem{JK.04}  V.E.\ Korepin,  Phys. Rev. Lett. {\bf 92} 096402 (2004);
B. -Q.\ Jin and V.E.\ Korepin, J.\ Stat.\ Phys.\ {\bf 116}, 79 (2004).
  \bibitem{LORV.05} J.I.\ Latorre,  R.\ Or\'us,
  E.\ Rico, and J. \ Vidal,  Phys. Rev. A {\bf 71}, 064101 (2005).
\bibitem{RCM.11}
R.\ Rossignoli, N.\ Canosa, and J.~M.\ Matera,
  Phys.\ Rev.\ A  {\bf 83}, 042328 (2011).
 \bibitem{BDV.06} T.\ Barthel, S.\ Dusuel, and J. \ Vidal,  Phys. Rev.
Lett. {\bf 97}, 220402 (2006);
J.\ Vidal, S.\ Dusuel, T.\ Barthel, J.\ Stat.\ Mech.:\ Theory Exp.\ (2007) P01015.
   \bibitem{WVB.10}H. Wichterich, J. Vidal, and S. Bose,
 Phys. Rev. A {\bf 81}, 032311 (2010).
\bibitem{MRC.082}
J.~M.\ Matera, R.\ Rossignoli, and N.\ Canosa,
  Phys.\ Rev.\ A {\bf 78}, 012316 (2008).
\bibitem{MRC.08}
J.~M.\ Matera, R.\ Rossignoli, and N.\ Canosa,
  Phys.\ Rev.\ A {\bf 78}, 042319 (2008).
 \end{thebibliography}
\end{document}